\documentclass[preprint,showkeys]{revtex4}
\usepackage{amssymb,amsmath}
\usepackage{epsfig}
\font\bba=msbm10 % scaled 1080
\font\bbb=msbm8 %scaled 1080
\font\bbc=msbm6 %scaled 1080
\newfam\bbfam
\textfont\bbfam=\bba
\scriptfont\bbfam=\bbb
\scriptscriptfont\bbfam=\bbc

\begin{document}
\title{Monte Carlo Simulations of the two-dimensional dipolar fluid}
\author{Jean-Michel Caillol}
\email{Jean-Michel.Caillol@th.u-psud.fr}
\affiliation{Univ. Paris-Sud, CNRS, LPT, UMR 8627, Orsay, F-91405, France}

\author{Jean-Jacques Weis}
\email{Jean-Jacques.Weis@th.u-psud.fr}
\affiliation{Univ. Paris-Sud, CNRS, LPT, UMR 8627, Orsay, F-91405, France}
\begin{abstract}
We study a two-dimensional fluid of dipolar hard disks by Monte Carlo simulations
in a square with periodic boundary conditions and on the surface of a sphere. The theory
of the dielectric constant and the asymptotic behaviour of the equilibrium pair correlation function
in the fluid phase is derived for both geometries. After having  established the equivalence of the 
two  methods we study the stability of the liquid phase in the canonical ensemble.
We give evidence of a phase made of living polymers at low temperatures and  provide a tentative  phase
diagram.
\end{abstract}
\keywords{
Two-dimensional dipolar fluid; Monte Carlo simulations; Periodic Boundary
conditions; Spherical boundary conditions.
}
\maketitle   
%%%%%%%%%%%%%%%%%%%%%%%%%%%%%%%%%%%%%%%%%%%%%%%%%%%%%%%
\section{Introduction}
This paper is devoted to a study of a two-dimensional (2D) system made of identical dipolar hard disks (DHD) in the Euclidian plane $E_2$
by means of Monte-Carlo (MC) simulations.
The dipoles are assumed to be permanent and the configurational energy of
$N$ dipolar molecules in  $E_2$ reads as
\begin{eqnarray}
\label{I1}
H= &&
\frac{1}{2}\sum_{i \ne j}^{N} v_{HS}(r_{ij}) 
+ \frac{1}{2} \mu^2 \sum_{i \ne j}^{N}
 \dfrac{1}{r_{ij}^2}  \left [ \mathbf{s}_i \cdot
\mathbf{s}_j - \frac{2(\mathbf{s}_i \cdot \mathbf{r}_{ij} )
 (\mathbf{s}_j \cdot \mathbf{r}_{ij} )}{ r_{ij}^{2}} \right ] \nonumber  \\  \; .
\end{eqnarray}
In Eq.\ \eqref{I1},  $v_{HS}(r)$ is the hard disk potential  of diameter $\sigma$.
The second term is the  contribution from the 2D dipole-dipole interaction
where $\boldsymbol{\mu}_i = \mu \mathbf{s}_i$, $\mu$ permanent dipole
moment, $\mathbf{s}_i$  unit vector in the direction of the
dipole moment of  particle $i$,
$\mathbf{r}_{ij}= \mathbf{r}_{j} - \mathbf{r}_{i}$, the vector joining the centres of mass 
of the particles, and $ r_{ij}= |\mathbf{r}_{ij}|$. 
We stress that the system that we  consider   cannot be
seen as a thin layer of  a real   3D  system of dipoles. In this  case, the electrostatic interactions 
should be derived from  the solutions of the  3D  Laplace equation while the dipole-dipole
interaction involved in Eq.~\eqref{I1} is derived from the solution of the 2D Laplace equation
in the plane.

We have performed MC simulations of the DHD fluid in a square with periodic boundary conditions  and on the surface
of an ordinary sphere. In both cases the dipole-dipole interaction is obtained from a rigorous solution
of Laplace equation in the considered geometry~\cite{perram:81,morris:85,Orsay_1,Caillol}. We compare the two methods in the liquid phase and check
that they both yield
the same thermodynamic, structural and dielectric properties. Both methods
are  then  used for preliminaries MC studies of the DHD fluid at low temperatures. In this regime, as for 
real  3D  dipoles confined in a plane (see \textit{e.g.}, Ref.~\cite{JJWeis_II}), the  2D  dipoles aggregate to form living
chains and ring polymers 
at low densities and more involved structures at higher densities. 

The paper is organized as follows. After this introduction  we give details on  the two  simulation
techniques used in this work in Sec.~\ref{Simu}. Next Sec.~\ref{dielectric} is devoted to  a digest of
the general theory of dielectric media in an arbitrary  2D  geometry~\cite{Caillol}, with applications to
the square with periodic boundary conditions and the sphere. 
This theoretical analysis
is notably  required to understand  the long range tails
of the pair correlation functions in both geometries. Checks of these asymptotic behaviours as well as
quantitative comparisons between the two methods are discussed in Sec.~\ref{compa}.
In Sec.~\ref{data} we present
extensive MC simulations of the DHD fluid by both methods and  give a tentative phase diagram
of the system. We conclude in Sec.~\ref{Conclu}

%%%%%%%%%%%%%%%%%%%%%%%%%%%%%%%%%
%%%%%%%%%%%%%%%%%%%%%%%%%%%%%%%%%
\section{Simulation methods}
\label{Simu}
%%%%%%%%%%%%%%%%%%%%%%%%%%%%%%%%%
\subsection{Periodic boundary conditions}
%%%%%%%%%%%%%%%%%%%%%%%%%%%%%%%%%
In this method the simulation cell is  a square of side $L$ with periodic
boundary conditions, that  will be referred to as space $\mathcal{C}_2$~\cite{perram:81,morris:85}.
Some care is required to take into account the long range
of dipole-dipole interaction. The usual way to compute the configurational energy
 $ U_{dd}$ is to replicate the basic simulation cell
periodically in space and calculate  $U_{dd}$ as the sum of
the interactions of the $N$ dipoles in the basic cell with all the other dipoles
in the cell and with the periodically repeated images in the surrounding
cells 
\begin{equation}
\label{II1}
 U_{dd} = \frac{\mu^2}{2} \sum_{i,j=1}^{N}  \sum_\mathbf{n} {}^{'} 
\biggl\{  \frac{ \mathbf{s}_{i} \cdot \mathbf{s}_{j} }
       {| \mathbf{r}_{ij} +L \mathbf{n}|^2}
   -2 \  \frac{ [\mathbf{s}_{i} \cdot (\mathbf{r}_{ij} +L \mathbf{n})]
             [\mathbf{s}_{j} \cdot (\mathbf{r}_{ij} + L \mathbf{n})]}
           {| \mathbf{r}_{ij} +L \mathbf{n}|^{4}}.
 \biggr\}
\end{equation}
The prime affixed to the sum over 
$\mathbf{n} = (n_x,n_y)$, with $n_x$,$n_y$ integers, means that the term $i
\neq j$  is omitted when $\mathbf{n}=0$.

 By a
lattice summation technique (Ewald sum) the slowly and conditionally  convergent sum is
transformed into two rapidly convergent sums, one in direct space, the
other in reciprocal space, the rate of convergence of both sums
being regulated by the parameter $\alpha$.
The resulting expression for the energy of the 2D system is~\cite{perram:81}
\begin{eqnarray}
\label{II11}
U_{dd}  =&& -\frac{\mu^2}{2} \sum_{i,j=1}^N  \sum_\mathbf{n} {}^{'}
  \left[b(|\mathbf{r}_{ij}+ L \mathbf{n}|) \mathbf{s}_i 
  \cdot\mathbf{s}_j \right. \nonumber \\ 
&&\left. +c(|\mathbf{r}_{ij} + L \mathbf{n}|)
  [\mathbf{s}_i \cdot 
  (\mathbf{r}_{ij} + L \mathbf{n})][\mathbf{s}_j \cdot (\mathbf{r}_{ij} 
+ L \mathbf{n})]\right] \nonumber \\
&& + \frac{\pi \mu^2}{S} \sum_{\mathbf{k} \neq 0} \displaystyle 
 \frac{ \exp (-k^2 /{4 \alpha^2})}{k^2}  F(\mathbf{k}) F^*(\mathbf{k}) \nonumber \\
&& - \alpha^2 \mu^2 \sum_{i=1}^N \mathbf{s}_i^2 +\displaystyle\frac{\pi \mu^2}{2 S} \displaystyle{ \left(1-
  \frac{\epsilon^{'} -1}{\epsilon^{'} +1}  \right)} \left(\sum_{i=1}^N 
   \mathbf{s}_i \right)^2  \; ,
\end{eqnarray}
where the functions $b(r)$ and $c(r)$ are given by
\begin{eqnarray}
\label{II12}
 b(r) &=&  - \frac{\exp(- \alpha^2 r^2)}{r^2}, \\
\label{II9}
 c(r) &=&    2 (\frac{1}{r^2} + \alpha^2) 
\frac{\exp(- \alpha^2 r^2)}{r^2},
\end{eqnarray}
and
\begin{eqnarray}
\label{II13}
F(\mathbf{k}) &=& \sum_{i=1}^N \left( \mathbf{k} \cdot \mathbf{s}_i
\right) \displaystyle \exp[i\mathbf{k} \cdot \mathbf{r}_i] \, .
\end{eqnarray}
In Eq.\ (\ref{II11}) $S=L^2$ is the
area of the simulation cell, $N$ the number of particles and $F^*$ the complex conjugate of $F$.
The wave-vectors $\mathbf{k}$ which enter the reciprocal space
contributions to the energy are of the form
\begin{equation}
\label{II10}
\mathbf{k} =2 \pi \mathbf{n}/L \;.
\end{equation}

Care has to be taken to properly choose the
$\alpha$ parameter which governs the rate of convergence of the
real- and reciprocal-space contributions in Eq.~\eqref{II11}. It is generally taken sufficiently
large so that only the terms with $\mathbf{n}=0$ need to be retained in 
Eqs.~ (\ref{II11}). 
The last term in Eq.~\eqref{II11} represents the contribution to the
energy from the depolarization field created  when a continuous medium
of dielectric constant $\epsilon'$ surrounds a disk shaped sample of
periodic replica. For a conducting medium ($\epsilon' = \infty$) this
term vanishes while for a system in vacuum ($\epsilon' = 1$) it is 
$\frac{\pi}{2S} \mathbf{M}^2$, where $\mathbf{M}$ is the total polarization of
the system.

A thermodynamic state of the DHD fluid is characterized by a reduced density $\rho^*=N\sigma^2/S$ where $S=L^2$
is the surface of the square of simulation and the  reduced dipole
$\mu^*$ with  $\mu^{*2}= \mu^2/(k_B T\sigma^2) $ ($k_B $ Boltzmann constant, $T$ temperature).
%%%%%%%%%%%%%%%%%%%%%%%%%%%%%%%%%
\subsection{Spherical  boundary conditions}
%%%%%%%%%%%%%%%%%%%%%%%%%%%%%%%%%
In this method the simulation cell is  the surface of an ordinary sphere of center $O$ and radius $R$,
that  will be referred to as space $\mathcal{S}_2$~\cite{Orsay_1,Caillol}.
The electrostatics can be solved exactly in   $\mathcal{S}_2$ in two different ways
and therefore two distinct models are  available~\cite{Trul_Cai,Caillol}.

In the first version  the DHD fluid is made of $N$ ordinary (or mono-) dipoles  $\boldsymbol{\mu}_i = \mu \mathbf{s}_i$
 tangent to the sphere  $\mathcal{S}_2$
at points $\mathbf{OM}_i=R \mathbf{z}_i$ ($\mathbf{z}_i \cdot \mathbf{s}_i =0$).
 In the second version considered
in this article, one rather considers a collection of $N$ bi-dipoles. A bi-dipole is defined
as a dumbell of two identical mono-dipoles located
at two antipodal points of the sphere at points
 $\mathbf{OM}_i=R \mathbf{z}_i$ and $\mathbf{O\overline{M}}_i= - R \mathbf{z}_i $. 
The numerical experiments of Ref.~\cite{Caillol} show that 
the convergence to the thermodynamic limit is in general faster for bi-dipoles  than 
for mono-dipoles. The configurational energy of the DHD fluid   reads 
\begin{equation}
\label{conf2}
 U(\{\mathbf{z}_i, \boldsymbol{\mu}_i \}) =  \frac{1}{2} \sum_{i \neq j}^N \; v_{\mathrm{HS}}^{\mathrm{bi}}(\psi_ {ij}) + \frac{1}{2}
\sum_{i \neq j}^N \;  W_{\boldsymbol{\mu}_i, \boldsymbol{\mu}_j}^{\mathrm{bi}}  \; ,
\end{equation}
where $v_{\mathrm{HS}}^{\mathrm{bi}}(\psi_ {ij}) $ is  hard-core pair potential  defined by
\begin{equation}
 v_{\mathrm{HS}}^{\mathrm{bi}}(\psi_ {ij}) =  \begin{cases}
                                                 \infty & \text{ if }  \sigma/R > \psi_{ij}  \text{ or }    \psi_{ij}> \pi -\sigma/R  \; ,   \\
                                                 0&       \text{ otherwise }  \; ,  
                                                \end{cases}
\end{equation}
where $\psi_ {ij}$ is the angle between vectors  $\mathbf{z}_i$ and  $\mathbf{z}_j$, \textit{i.e.} $\cos \psi_ {ij}=\mathbf{z}_i , \cdot \mathbf{z}_j $
and thus $r_{ij}= R \psi_ {ij}$ is the length of the geodesic length between points $M_i$
and $M_j$.
The dipole-dipole interaction $W_{\boldsymbol{\mu}_i, \boldsymbol{\mu}_j}^{\mathrm{bi}} $ is given by
\begin{equation}
\label{ene_bi}
 W_{\boldsymbol{\mu}_i ,\boldsymbol{\mu}_j}^{\mathrm{bi}} =  \frac{\mu^2}{R^2}\frac{1}{\sin^2\psi_{ij}} \bigg(   \mathbf{s}_i  \cdot  \mathbf{s}_j  
+  \frac{2 \cos \psi_{ij}}{\sin^2 \psi_{ij}}\, (   \mathbf{s}_i \cdot \mathbf{z}_j)
 ( \mathbf{s}_j \cdot\mathbf{z}_i)
 \bigg) \; .
\end{equation}

In Eq.~\eqref{conf2} the vectors $\mathbf{z}_i$ can always be chosen
in the northern hemisphere  $\mathcal{S}_2^+$ because of the special symmetries of the interaction.
It is thus clear that the actual  domain
occupied by the fluid is  the northern hemisphere  $\mathcal{S}_2^+$ rather than the whole
hypersphere.
In terms of mono-dipoles the interpretation of the model is therefore the  following :
when a mono-dipole $\boldsymbol{\mu}_i$ leaves $\mathcal{S}_2^+$ at some point $M_i$ of the equator
the same dipole moment  $\boldsymbol{\mu}_i$  reenters $\mathcal{S}_2^+$ at the antipodal point  $\overline{M}_i$.
Therefore bi-dipoles living
on the whole sphere are equivalent to mono-dipoles living on the northern hemisphere but with
special  boundary conditions ensuring homogeneity and isotropy at equilibrium (in the case of a fluid phase). 
We stress that the expression~\eqref{ene_bi} has been deduced rigorously from the solution of Laplace-Betrami equation in 
 $\mathcal{S}_2$~\cite{Caillol} by contrast with the heuristic dipole-dipole interaction used in reference~\cite{Orsay_1}.

A thermodynamic state of this model is now characterized by a dimensionless number density $\rho^*=N\sigma^2/S$ where $S=2 \pi R^2$
is the $2 D$ surface  of the northern hemisphere  $\mathcal{S}_2$ and the  reduced dipole
$\mu^*$ with  $\mu^{*2}= \mu^2/(k_B T\sigma^2) $. 
%%%%%%%%%%%%%%%%%%%%%%%%%%%%%%%%%%%%%%
%%%%%%%%%%%%%%%%%%%%%%%%%%%%%%%%%%%%%%
\section{Fulton's theory}
\label{dielectric}
Let us  consider quite generally a polar fluid occupying  a  2D  surface $\Lambda$ with boundaries $\partial \Lambda$.
We assume the system to be at thermal equilibrium in a homogeneous and isotropic fluid phase. The fluid behaves
macroscopically as a dielectric medium characterized by a scalar dielectric constant $\epsilon$.
Due to the lack of screening in such fluids, the asymptotic behaviour of the pair correlation function is long ranged and
depends on the geometry of the system, \textit{i.e.} its shape, size, and the  properties imposed to
the electric field (or potential) on the boundaries 
$\partial \Lambda$ as well. 
As a consequence, the expression of the dielectric constant $\epsilon$ in terms of the fluctuations of polarization
also depends on the geometry.
These issues can  be formally taken into account in the framework of Fulton's theory~\cite{Fulton_I,Fulton_II,Fulton_III}
which achieves
an elegant  synthesis between  the linear response theory  and 
the electrostatics of continuous media.

In addition to provide an expression for the dielectric constant $\epsilon$ Fulton's formalism also yields
the asymptotic behaviour of the pair correlation function.
Fulton's formalism can be extended without more ado to non-euclidian geometries and was applied
notably  to  3D   cubic systems with periodic boundary conditions (space $\mathcal{C}_3$) and hyperspheres $\mathcal{S}_3$
in Refs.~\cite{Caillol_4,Trul_Cai},
and, recently,  to the  2D  euclidian plane $E_2$ and the sphere $\mathcal{S}_2$, for both
mono- and bi-dipoles~\cite{Caillol}. In this section we derive the missing results for space $\mathcal{C}_2$
and recall the results for  2D  polar fluids in $E_2$ and $\mathcal{S}_2$.

Fulton's relations constitute  the quintessence of Fulton's formalism; they are  formally independent of the geometry and  read
\begin{subequations}\label{F_rel}
 \begin{align}
\boldsymbol{\chi} =&  \boldsymbol{\sigma}  +  \boldsymbol{\sigma} \circ  \mathbf{G} \circ \boldsymbol{\sigma}  \label{F_rel_a} \\ 
 \mathbf{G}          =&    \mathbf{G}_0  \circ ( \mathbf{I} -  \boldsymbol{\sigma}  \circ \mathbf{G}_0)^{-1}     \label{F_rel_b}                                                
 \end{align}
\end{subequations}
Some comments seem appropriate. Let us first define the tensorial susceptibility  $\boldsymbol{\chi} $.
Under the influence of  an external electrostatic field $\boldsymbol{\mathcal{E}}(\mathbf{r}) $
the medium  acquires a macroscopic polarization 
\begin{equation}
 \mathbf{P} (\mathbf{r}) = < \widehat{\mathbf{P}} (\mathbf{r})>_{\boldsymbol{\mathcal{E}}}  \; ,
\end{equation}
where the brackets denote the equilibrium average of the microscopic polarization
$\widehat{\mathbf{P}} (\mathbf{r}) = \sum_{j=1}^N  
\boldsymbol{\mu}_j \, \delta^{(2)}( \mathbf{r}- \mathbf{r}_j)$ in the presence of the external field 
 $\boldsymbol{\mathcal{E}}$. 
The relation between the macroscopic polarization $\mathbf{P}$ and the external field 
 $\boldsymbol{\mathcal{E}} $ can be established in the framework of
linear-response theory, provided that  $\boldsymbol{\mathcal{E}}$
is small enough,  with the result
\begin{equation}
\label{LRT}
 2 \pi \mathbf{P}(\mathbf{r}_1) =\left[\boldsymbol{\chi} \circ  \boldsymbol{\mathcal{E}}\right](\mathbf{r}_1)
 \left(\equiv 
 \int_{\Lambda}  d^{2} \mathbf{r}_2 \;\boldsymbol{\chi} (\mathbf{r}_1,\mathbf{r}_2)  \cdot \boldsymbol{\mathcal{E}}(\mathbf{r}_2) \right) \; .
\end{equation}
The r.h.s. of Eq.~\eqref{LRT} has been formulated in  a compact, albeit convenient notation that will
be adopted henceforth,  where
the symbol $\circ$ (which also enters Eqs.~\eqref{F_rel}) means both a tensorial contraction (denoted by the dot '' $\cdot $ '')
and a spacial convolution over the whole domain $\Lambda$ filled by the medium.
The  tensorial susceptibility  $\boldsymbol{\chi} $ in Eq.~\eqref{F_rel_a} reads
\begin{equation}
\label{chi}
\boldsymbol{\chi} (\mathbf{r}_1, \mathbf{r}_2) = 2 \pi \beta <\widehat{\mathbf{P}} (\mathbf{r}_1) \widehat{\mathbf{P}} (\mathbf{r}_2)> \; ,
\end{equation}
where $\beta = 1/k_BT$ and the thermal averages are computed in the absence of the external field $\boldsymbol{\mathcal{E}}$.

However, the dielectric properties of the fluid are characterized by the dielectric tensor $\boldsymbol{\epsilon}$. In Eqs~\eqref{F_rel}
we have introduced, following Fulton, the convenient notation $\boldsymbol{\sigma}=\boldsymbol{\epsilon} - \mathbf{I} $
with $\mathbf{I}(\mathbf{r}_1, \mathbf{r}_2)=\mathbf{U} \delta^{(2)}(\mathbf{r}_{12})$
where  $\mathbf{U}= \mathbf{e}_x \mathbf{e}_x + \mathbf{e}_y\mathbf{e}_y$ is  the unit dyadic tensor.
The tensor $\boldsymbol{\sigma}$ enters the constitutive  relation
\begin{equation}
\label{Max}
  2 \pi \mathbf{P}= \boldsymbol{\sigma} \circ \mathbf{E}  \; ,
\end{equation}
where the Maxwell field $\mathbf{E}(\mathbf{r})$ is the sum of the  external field 
$\boldsymbol{\mathcal{E}}(\mathbf{r}) $ and the electric field created  by the macroscopic polarization
of the fluid.  Therefore one has 
\begin{equation}
\label{zon}
 \mathbf{E} =  \boldsymbol{\mathcal{E}} + 2 \pi \mathbf{G}_0 \circ \mathbf{P} \; ,
\end{equation}
where $\mathbf{G}_0$ denotes the bare dipolar Green's function.  Note that
$2 \pi \mathbf{G}_0 (\mathbf{r}_1,\mathbf{r}_2) \cdot \boldsymbol{\mu}_2$ is the electric field
at point $\mathbf{r}_1$ created by a point dipole $ \boldsymbol{\mu}_2$ located at point $\mathbf{r}_2$
in vacuum and in the presence of the boundary $\partial \Lambda$. In the presence of
the dielectric medium this field is now given by 
$2 \pi \mathbf{G}_ (\mathbf{r}_1,\mathbf{r}_2) \cdot \boldsymbol{\mu}_2$
where the macroscopic, or dressed, Green's function is given by Eq.~\eqref{F_rel_b}  in 
which the inverse must be understood in the sense of operators. 

It is generally assumed that $\boldsymbol{\epsilon}$ is a local function,
\textit{i.e.}  $\boldsymbol{\epsilon}= \epsilon \mathbf{I}$. 
More precisely, it is plausible -and we shall take it for granted- that 
$\boldsymbol{\epsilon}(\mathbf{r}_1,\mathbf{r}_2)$
is a short range function of the distance between the two points $\mathbf{r}_1$ and $\mathbf{r}_2$, 
at least for a homogeneous liquid (or in the bulk in the presence of interfaces),
and one then  defines 
\begin{equation}
 \epsilon \mathbf{U} =
 \int_{\Lambda}   d^{2} \mathbf{r}_2   \;  \boldsymbol{\epsilon}(\mathbf{r}_1,\mathbf{r}_2) \; .
\end{equation}
Experiments show, and this fact must be admitted, that while  $\boldsymbol{\epsilon}$ is an intrinsic
property of the medium Eqs.~\eqref{F_rel} show that the susceptibility tensor
$\boldsymbol{\chi} (\mathbf{r}_1,\mathbf{r}_2)$ is a long range function of $\mathbf{r}_{12}$ which 
depends on the considered geometry.  The locality assumption on  $\boldsymbol{\epsilon}$
allows an explicit calculation of the Green's function $ \mathbf{G}(\mathbf{r}_1,\mathbf{r}_2)$ in some
geometries, notably those used in MC simulations.

%%%%%%%%%%%%%%%%%%%%%%%%%%%%%%%%%%%%%%
%%%%%%%%%%%%%%%%%%%%%%%%%%%%%%%%%%%%%%
\subsection{The square $\mathcal{C}_2$}
In Ref.~\cite{Caillol_4}  Fulton's formalism was applied to the  3D  dipolar hard sphere fluid in $\mathcal{C}_3$.
Here we will consider the   2D  case which is a mere transposition so that we can skip many details. Moreover
we will consider only tinfoil boundary conditions, \textit{i.e.} $\epsilon^{'}= \infty$, which simplifies the algebra.
The bare Green's function is given by
\begin{equation}
 \mathbf{G}_0(\mathbf{r}_1, \mathbf{r}_1) \equiv \mathbf{G}_0(\mathbf{r}_{12}) =
  \dfrac{1}{2 \pi}\dfrac{\partial}{\partial \mathbf{r}_{12}} 
\dfrac{\partial}{\partial \mathbf{r}_{12}}  \psi(\mathbf{r}_{12}) \; ,
\end{equation}
where  $\psi(\mathbf{r})$ is the periodic Ewald potential. It satisfies Poisson's equation in  $\mathcal{C}_2$
\begin{equation}
 \Delta \psi(\mathbf{r}) = -2 \pi [\delta_{\mathcal{C}_2}  ( \mathbf{r})     - \dfrac{1}{L^2} ] \; ,
\end{equation}
where 
\begin{align}
 \delta_{\mathcal{C}_2}  ( \mathbf{r})= &\sum_{\mathbf{n}} \delta^{(2)}(\mathbf{r} -L \mathbf{n}) \nonumber \\
=& \dfrac{1}{L^2}\sum_{\mathbf{k}} \exp (i\mathbf{k} \cdot  \mathbf{r})  \, ,
\end{align}
is the periodical Dirac's comb.  Expanding  $\psi(\mathbf{r})$ and 
$\mathbf{G}_0(\mathbf{r})$ in Fourier series
one finds
\begin{subequations}
 \begin{align}
 \psi(\mathbf{r}) =& \dfrac{2 \pi}{L^2} \sum_{\mathbf{k} \neq \mathbf{0} }\dfrac{\exp(i\mathbf{k} \cdot  \mathbf{r})}{\mathbf{k}^2} \\
 \mathbf{G}_0(\mathbf{r}) =& - \dfrac{1}{L^2} \sum_{\mathbf{k} \neq \mathbf{0}} \widehat{\mathbf{k}} \widehat{\mathbf{k}}   \exp(i\mathbf{k} \cdot  \mathbf{r}) \; ,
 \end{align}
\end{subequations}
where  $\widehat{\mathbf{k}} =  \mathbf{k}/\Arrowvert  \mathbf{k}  \Arrowvert$.

It shall proof useful in what follows  to remark that one can rewrite  the Ewald potential as
\begin{equation}
 \psi(\mathbf{r}) = - \log r + \dfrac{\pi}{2 L^2} r^2 + \delta  \psi(\mathbf{r}) \; ,
\end{equation}
where $\delta  \psi(\mathbf{r}) $ is a harmonic function  in the square and can thus be expressed quite
generally as~\cite{Jackson,Felderhof}
\begin{equation}
 \delta  \psi(\mathbf{r}) = \sum_{m=1}^{\infty} a_m r^m \cos(m \varphi + \alpha_m) \; ,
\end{equation}
where $\varphi$ is the angle of $\mathbf{r}$ with the axis  $\mathbf{e}_x$  and   the constants $a_m$, $\alpha_m$ 
are such that  $\psi(\mathbf{r}) $ is a periodical function.

It is then easy to deduce from these prolegomena the two formulas
\begin{subequations}\label{G0}
  \begin{align}
   \mathrm{Tr} \,  \mathbf{G}_0(\mathbf{r}_1, \mathbf{r}_2) =&\dfrac{1}{L^2} - \delta_{\mathcal{C}_2}  ( \mathbf{r}_{12})
\label{G0_a} \; , \\
2 \widehat{\mathbf{r}}_{12} \cdot  \,  \mathbf{G}_0(\mathbf{r}_1, \mathbf{r}_2)
\cdot  \widehat{\mathbf{r}}_{12}   =&  \dfrac{1}{L^2} + \dfrac{1}{\pi r_{12}^2} + 
                                                      \dfrac{1}{\pi}\sum_{m=2}^{\infty} m(m-1)a_m r^{m-2} \cos(m \varphi_{12} + \alpha_m) \; .
\label{G0_b}
  \end{align}
\end{subequations}

The computation of the dressed Green's function from its definition~\eqref{F_rel_b} is conveniently 
made in Fourier space.
Under the assumption of the locality of the dielectric tensor $\boldsymbol{\epsilon}(\mathbf{r}_1, \mathbf{r}_2)$
one finds the obvious result  $\mathbf{G}=\mathbf{G}_0/\epsilon$.  Therefore Fulton's relation~\eqref{F_rel_a} takes  the explicit
form
\begin{equation}
\label{Fulton}
\boldsymbol{\chi}(\mathbf{r}_1, \mathbf{r}_2) =
(\epsilon-1) \mathbf{I} (\mathbf{r}_1, \mathbf{r}_2) +
\dfrac{(\epsilon-1)^2}{\epsilon}  \mathbf{G}_0(\mathbf{r}_1, \mathbf{r}_2) \, .
\end{equation}
We stress that the above equation has been obtained under the assumption of the locality
of the dielectric tensor  $\boldsymbol{\epsilon}(\mathbf{r}_1,\mathbf{r}_2)$. Therefore it should be valid only asymptotically,
\textit{i.e.} for points $(\mathbf{r}_1,\mathbf{r}_2)$ at a mutual distance $r_{12}$ larger then the range 
$\xi$ of $\boldsymbol{\epsilon}(\mathbf{r}_1,\mathbf{r}_2)$.

Taking the trace of Eq.~\eqref{Fulton}, making use of Eq.~\eqref{G0_a} and  integrating both $\mathbf{r}_1$ and $\mathbf{r}_2$ over
the square $\mathcal{C}_2 $ one finds the expression of the dielectric constant
\begin{equation}
\label{formu_eps}
 \epsilon -1 = \dfrac{\pi \beta}{L^2} <\mathbf{M}^2> \; ,
\end{equation}
where $\mathbf{M}= \sum_{i=1}^{N} \boldsymbol{\mu}_i$ is the total dipole moment of the square.

We turn now our attention to the susceptibility tensor $\boldsymbol{\chi}(\mathbf{r}_1, \mathbf{r}_2)$
which may be expressed  in terms of the pair correlation function
$g(1,2)$ where $i \equiv (\mathbf{r}_i, \alpha_i)$ ($i=1,2$) denotes the position and the angle of dipole
 $\boldsymbol{\mu}_i$ with
 axis $\mathbf{e}_x$. One obtains that
\begin{equation}
\label{chi_b}
 \boldsymbol{\chi}(\mathbf{r}_1, \mathbf{r}_2) = y \mathbf{I} (\mathbf{r}_1, \mathbf{r}_2)+
2 y \rho \int_0^{2 \pi} \dfrac{d \alpha_1}{2 \pi}\int_0^{2 \pi} \dfrac{d \alpha_2}{2 \pi} \; h(1,2) \mathbf{s}_1 \mathbf{s}_2 
\; ,
\end{equation}
where $y=\pi \beta \rho \mu^2$ and $h=g-1$ as usual. In the infinite plane $E_2$ the
pair correlation function $g(1,2) $ can be expanded on a complete set of rotational invariants
among which  the most important  are
\begin{subequations}
 \begin{align}
\Phi_{00}(1,2) = & 1           \; , \\
  \Delta(1,2) = & \mathbf{s}_1 \cdot \mathbf{s}_2   \; ,\\
   D(1,2) = & 2(\mathbf{s}_1 \cdot \widehat{\mathbf{r}}_{12}) (\mathbf{s}_2 \cdot \widehat{\mathbf{r}}_{12})  -  \Delta(1,2) \; ,
 \end{align}
\end{subequations}
where $\widehat{\mathbf{r}}_{12} = \mathbf{r}_{12} /r_{12}$.

In space  $\mathcal{C}_2$ the function $g(1,2)$ has the symmetry of the square and \textit{stricto sensu}
cannot be expanded
onto these rotational invariants. However, following de Leeuw \textit{et al.}\cite{Deleeuw} one  defines
the projections
\begin{subequations}
\label{titi}
  \begin{align}
   h^{\Delta}(\mathbf{r}_{12}) =& 2   \int_0^{2 \pi}\dfrac{d \alpha_1}{2 \pi}  \int_0^{2 \pi} \dfrac{d \alpha_2}{2 \pi}\; h(1,2)   \Delta(1,2) \; , \\
   h^{D}(\mathbf{r}_{12}) =& 2   \int_0^{2 \pi}\dfrac{d \alpha_1}{2 \pi}  \int_0^{2 \pi} \dfrac{d \alpha_2}{2 \pi}\; h(1,2)   D(1,2) \; . \\
  \end{align}
\end{subequations}
Note that the two projections   $h^{\Delta}(\mathbf{r}_{12}) $ and  $h^{D}(\mathbf{r}_{12}) $ are
periodic functions which depend explicitely on the direction of vector $ \mathbf{r}_{12}$. The susceptibility
tensor  $\boldsymbol{\chi}(\mathbf{r}_1, \mathbf{r}_2)$ cannot be expressed in terms of these
sole projections; however one can deduce from Eq.~\eqref{chi_b} and the definitions~\eqref{titi} the relations
\begin{subequations} \label{Tr}
  \begin{align}
   \mathrm{Tr}\boldsymbol{\chi}(\mathbf{r}_1, \mathbf{r}_2) =& 2 y \;  \delta_{\mathcal{C}_2}  ( \mathbf{r}_{12}) + y \rho   h^{\Delta}(\mathbf{r}_{12}) \; , \\
2 \widehat{\mathbf{r}}_{12} \cdot  \,\boldsymbol{\chi} (\mathbf{r}_1, \mathbf{r}_2)
\cdot  \widehat{\mathbf{r}}_{12}  =& y \rho   h^{D}(\mathbf{r}_{12})  \; .
  \end{align}
\end{subequations}
The comparison of Eqs.~\eqref{G0},~\eqref{Tr}, and ~\eqref{Fulton} yields the asymptotic behaviour of
the projections  $h^{\Delta}(\mathbf{r}_{12}) $ and  $h^{D}(\mathbf{r}_{12}) $, \textit{i.e.}, for
 $\Arrowvert \mathbf{r}_{12}\Arrowvert > \xi$. One has
\begin{subequations}
\label{asymp_a}
  \begin{align}
    h^{\Delta}_{\text{asymp}}(\mathbf{r}_{12}) =& \dfrac{(\epsilon-1)^2}{\epsilon} \dfrac{1}{y \rho} \dfrac{1}{L^2} \; , \\
 h^{D}_{\text{asymp}}(\mathbf{r}_{12}) =& \dfrac{(\epsilon-1)^2}{\epsilon}\dfrac{1}{y \rho \pi}  \biggl\{
\dfrac{1}{ r_{12}^2}
+ \sum_{m=2}^{\infty}m(m-1) a_m r_{12}^{m-1}\cos(m \varphi_{12} + \alpha_m) 
 \biggr\} \; .
  \end{align}
\end{subequations}
In actual simulations one rather computes  angular averages of the functions   $h^{\Delta}(\mathbf{r}_{12})$
and $h^{D}(\mathbf{r}_{12})$, \textit{i.e.},

\begin{equation}
 h^{\Delta (D) }(r_{12}) = \int_0^{2 \pi} \dfrac{d \varphi_{12}}{2 \pi} h^{\Delta(D)}(\mathbf{r}_{12}) \; .
\end{equation}
The asymptotic values of these averaged functions are simpler and given by
\begin{subequations}
 \label{asymp_b}
  \begin{align}
   \label{trombo}     h^{\Delta}_{\text{asymp}}(r_{12}) =&  \dfrac{(\epsilon-1)^2}{\epsilon} \dfrac{1}{y \rho} \dfrac{1}{L^2} \; , \\
    h^{D}_{\text{asymp}}(r_{12}) =& \dfrac{(\epsilon-1)^2}{\epsilon}  \dfrac{1}{y \rho} \dfrac{1}{\pi r_{12}^2} \; ,
  \end{align}
\end{subequations}
which are  valid of course only  for $\xi < r_{12} < L/2$.
We  note that, in the thermodynamic
limit : \textit{i.e.} for $r$ fixed and $L \to \infty$,  one recovers the expected
Euclidian behaviours     $h^{\Delta}_{\text{asymp}}(r)\sim 0$ (\textit{i.e.} a short range function of $r$)  
and    $ h^{D}_{\text{asymp}}(r) \sim (\epsilon-1)^2/( \pi y \rho \epsilon) \times 1/r^2$ valid for
the Euclidian  plane $E_2$  without boundaries at infinity (cf. Refs.~\cite{Orsay_1,Caillol}).

Our last comment concerns Eq.~\eqref{formu_eps} which can be recast as
\begin{equation}
 \epsilon -1 =y \biggl\{ 1 + \dfrac{\rho}{2} \int_{\mathcal{C}_2}
d^2 \mathbf{r} \; h^{\Delta}(\mathbf{r})\biggl\}
\end{equation}
that we examine in the  limit $L \to \infty$. We  can then write
\begin{equation}
\label{chto}
  \epsilon -1 = \biggl\{ 1 + \dfrac{\rho}{2} \int_{E_2}
d^2 \mathbf{r} \; h^{\Delta}_{\infty}(r)\biggl\} + \dfrac{\rho}{2}\int_{\mathcal{C}_2} d^2 \mathbf{r} \; 
h^{\Delta}_{\text{asymp}}(r) \; ,
\end{equation}
where we have noted that, in the  limit $L \to \infty$,  $h^{\Delta}(\mathbf{r}) \to h^{\Delta}_{\infty}(r)$ 
becomes an isotropic
function. Making use of Eq.~\eqref{trombo} to compute the second integral in~\eqref{chto} one  obtains
\begin{equation}
\label{eps_E2}
 \dfrac{( \epsilon -1)( \epsilon +1)  }{2 \epsilon} = y \biggl\{ 1 + \dfrac{\rho}{2} \int_{E_2}
d^2 \mathbf{r} \; h^{\Delta}_{\infty}(r)\biggl\} \; .
\end{equation}
This expression of  $\epsilon$ is precisely that obtained  in space $E_2$ by
various methods~\cite{Orsay_1,Caillol}.

%%%%%%%%%%%%%%%%%%%%%%%%%%%%%%%%%%%%%%
%%%%%%%%%%%%%%%%%%%%%%%%%%%%%%%%%%%%%%
\subsection{The sphere $\mathcal{S}_2$}

We recall here the results of Ref.~\cite{Caillol} for a fluid of bi-dipoles confined on the
surface of the  sphere $\mathcal{S}_2$.
The dielectric constant is given by
  \begin{equation}
\label{eps_C2}
   \frac{\epsilon -1}{\epsilon} +  \frac{(\epsilon -1)^2}{2 \epsilon} \, \cos \psi_0 = \mathbf{m}^2(\psi_0) 
    \; 
 \text{  with  }   0<  \psi_0 <\pi/2 \; ,
  \end{equation}
where the fluctuation $\mathbf{m}^2(\psi_0)$ is  given by
  \begin{equation}
   \mathbf{m}^2(\psi_0) = \frac{ \pi \beta \mu^2}{S}  < \sum_i^N \sum_j^N \mathbf{s}_i \cdot  \mathbf{s}_j \, \Theta (\psi_0 -\psi_{ij})> \; ,
  \end{equation}
with $S=2 \pi R^2$ (surface of the northern hemisphere) and  $\Theta(x)$  the Heaviside step-function ($\Theta(x)=0$ for $x<0$ and  $\Theta(x)=1$ for $x>0$).
In the MC simulations reported in this paper we retained the optimal choice $\psi_0 =\pi /3$.
Asymptotically
(\textit{i.e.}  for a large fixed $r=R \psi \gg \xi $ and   $\psi < \pi/2$),  one has
\begin{subequations}
\label{assym_bi}
  \begin{align}
   h^{\Delta}_{\text{asymp}}(r) &  \sim  -\frac{(\epsilon-1)^2}{y \rho \epsilon} \frac{1}{2 \pi R^2 }  \frac{1}{1  +  \cos \psi}   \; , \\
   h^{D}_{\text{asymp}}(r)          & \sim \frac{(\epsilon-1)^2}{y \rho \epsilon} \frac{1}{2 \pi R^2 }\frac{1}{1  -  \cos \psi} \; .
 \end{align}
 \end{subequations}
As for $\mathcal{C}_2$ these asymptotic behaviours allow to recover from the formula~\eqref{eps_C2} of the 
dielectric constant in space $\mathcal{S}_2$ the expression~\eqref{eps_E2} in the thermodynamic limit.

%%%%%%%%%%%%%%%%%%%%%%%%%%%%%%%%%%%%%%
%%%%%%%%%%%%%%%%%%%%%%%%%%%%%%%%%%%%%%
\section{Comparisons of the two geometries of simulation}
\label{compa}
We performed standard MC simulations of the DHD fluid in the canonical ensemble
with single particle displacement moves (translation and rotation) in both geometries
$\mathcal{C}_2$ and $\mathcal{S}_2$. Some elements of comparison are given
in  Table~\ref{I} for three equilibrium states in the isotropic fluid phase of the model. 
We report values for the reduced internal energy per particle  $\beta u=<\beta U_{dd}>/N$, 
the contact values of the projections
 $g^{00}(\sigma)$,    $h^{\Delta}(\sigma)$, and  $h^{D }(\sigma)$ of the pair correlation function $g(1,2)$, the compressibility factor
 $Z=\beta P / \rho$ ( $P$ the pressure) with $Z=Z_{HS}+\beta u$ and $Z_{HS}=1 + (\pi \rho^*/2)g^{00}(\sigma) $,
and the  specific heat $C_v/k_B=(<(\beta U_{dd})^2> -<\beta U_{dd}>^2)/N$.
As apparent in  Table~\ref{I}, the agreement between the two methods of simulation is quite satisfactory.
 The values
reported in the table  were obtained for systems of $N \sim 1000$ particles for which   $N_{\rm Conf.}=5-10 \times 10^6$ configurations
per particle  were generated.
Note that the  finite size scaling study of Ref.~\cite{Caillol} gives, in the thermodynamic limit $N \to \infty$, $\beta u_{\infty} =-1.79000(6)$ 
for state ($\rho^{*}=0.7$, $\mu^*=\sqrt{2}$) and $\beta u_{\infty} = -4.17138(15)$ for state ($\rho^{*}=0.6$, $\mu^*=2$), which shows
that the data reported here are not very far from this limit. 

\begin{figure}[h!]
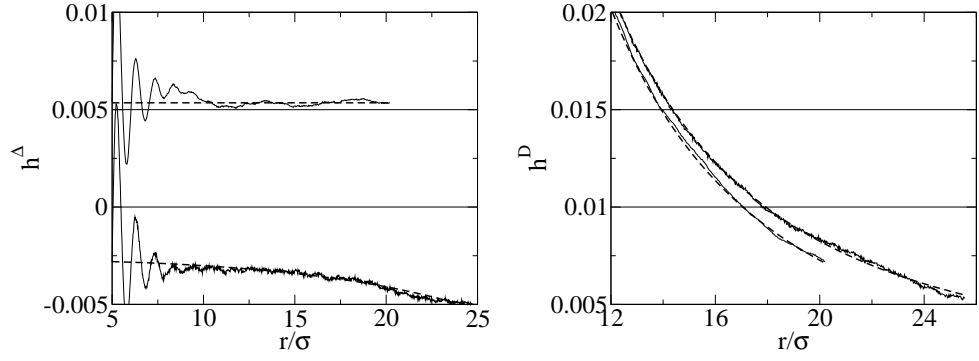

\centering
\begin{tabular}{ c c c }
\vspace{1.em} \\
\includegraphics[scale=0.3]{h_Delta.eps} & \hspace{0.2em} &  \includegraphics[scale=0.3]{h_D.eps}
\end{tabular}
\caption{Projections   $h^{\Delta}$ (Top : $\mathcal{C}_2$, bottom $\mathcal{S}_2$)
 and $h^{D}$ (Top : $\mathcal{S}_2$, bottom $\mathcal{C}_2$)  for the state ($\rho^{*} =0.6$, $\mu^* = 2$).
Solid lines : MC data, dashed lines : predicted asymptotic behaviours. In space $\mathcal{C}_2$ : $r<L/2$ while
in space $\mathcal{S}_2$ : $r<R \pi/2$.}
\label{Fig1}
\end{figure}

We have also tested the validity of the asymptotic behaviours of $h^{\Delta}(r)$, and $h^{D}(r)$ in both geometries.
We display in Fig.~\ref{Fig1} these functions as well as their asymptotic behaviours~\eqref{asymp_b} and~\eqref{assym_bi}
for the state ($\rho^{*} =0.6$, $\mu^* = 2$). The values of the dielectric constant which enter these asymptotic behaviours
are those given in  Table.~\ref{I}.  As apparent on the figures an excellent agreement between the MC data and the theoretical prediction
is obtained.  The small tails observed in  $h^{\Delta}(r)$ at large $r$,  which differ significantly in the two geometries,
are of primary importance to ensure that the dielectric constants $\epsilon$ are identical in both geometries, 
within numerical uncertainties and finite size effects,
although given by completely different formulas.

%%%%%%%%%%%%%%%%%%%%%%%%%%%%%%%%%%%%%%
%%%%%%%%%%%%%%%%%%%%%%%%%%%%%%%%%%%%%%
\section{MC simulations of the fluid phase}
\label{data}
%%%%%%%%%%%%%%%%%%%%%%%%%%%%%%%%%%%%%%
%%%%%%%%%%%%%%%%%%%%%%%%%%%%%%%%%%%%%%

\begin{figure}[h!]
\centering
\begin{tabular}{ c c c }
\includegraphics[scale=0.4]{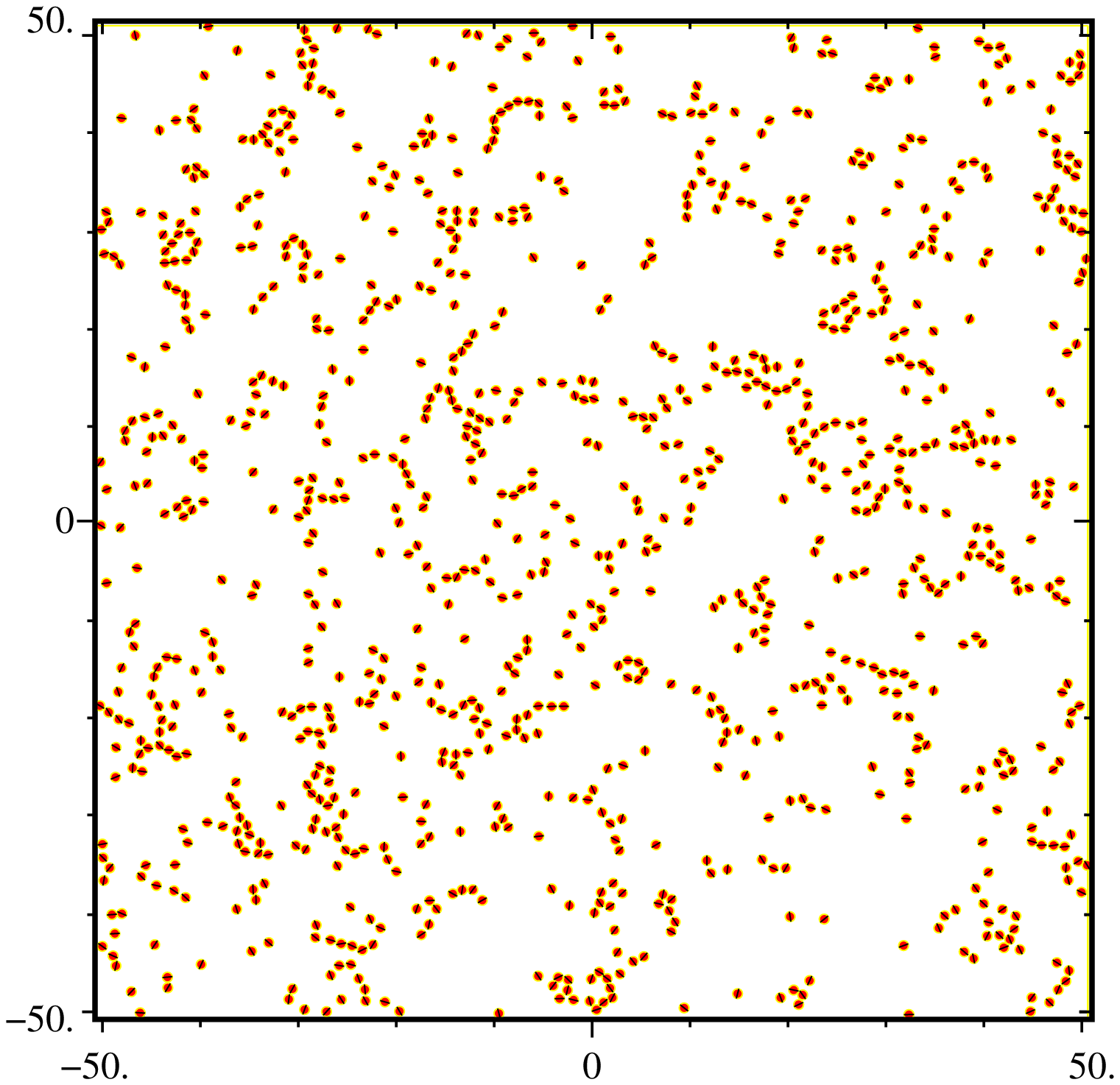} & \hspace{0.2em} &  \includegraphics[scale=0.4]{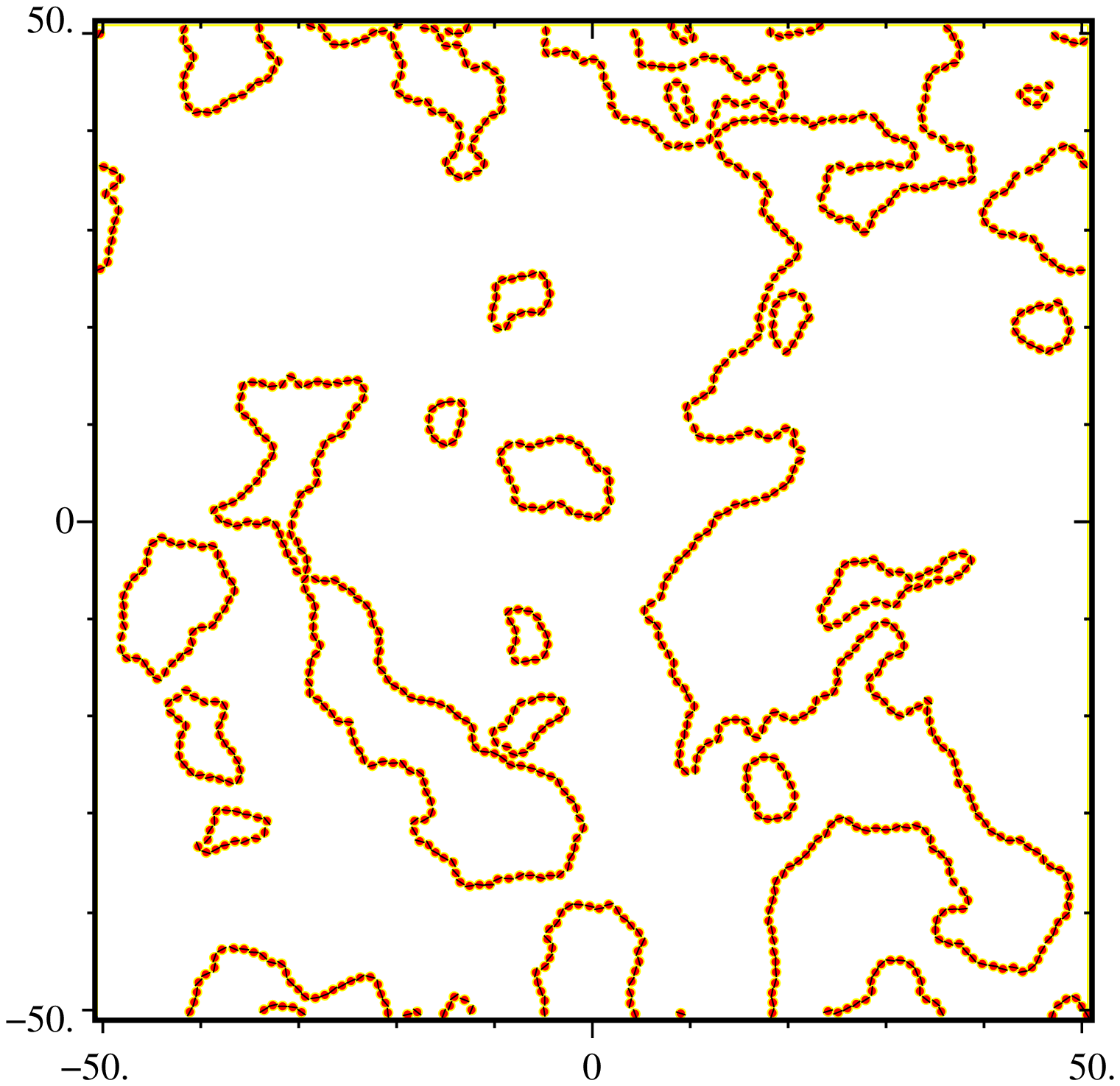} \\
\vspace{1.em} \\
\includegraphics[scale=0.4]{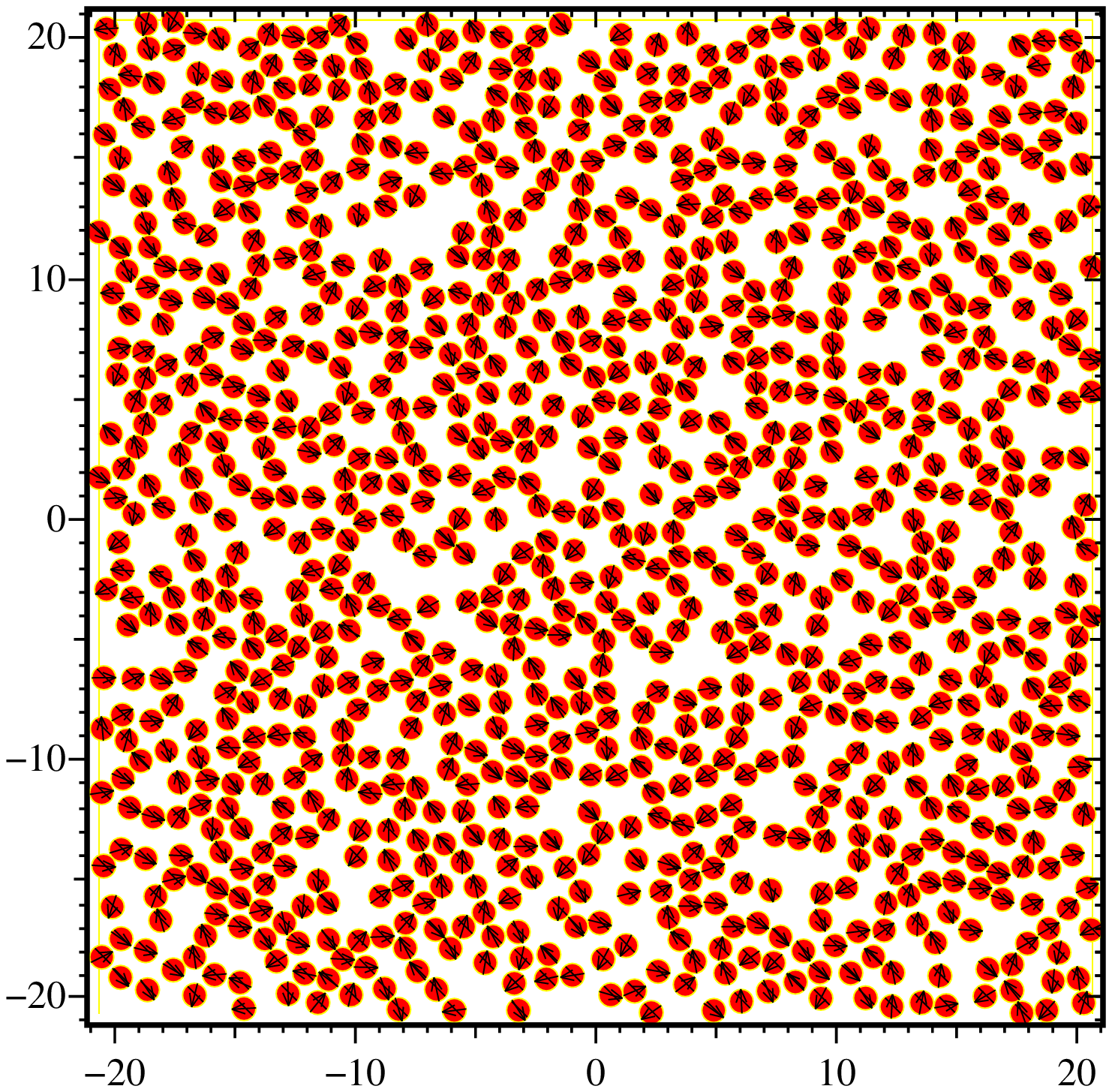} & \hspace{0.2em} &  \includegraphics[scale=0.4]{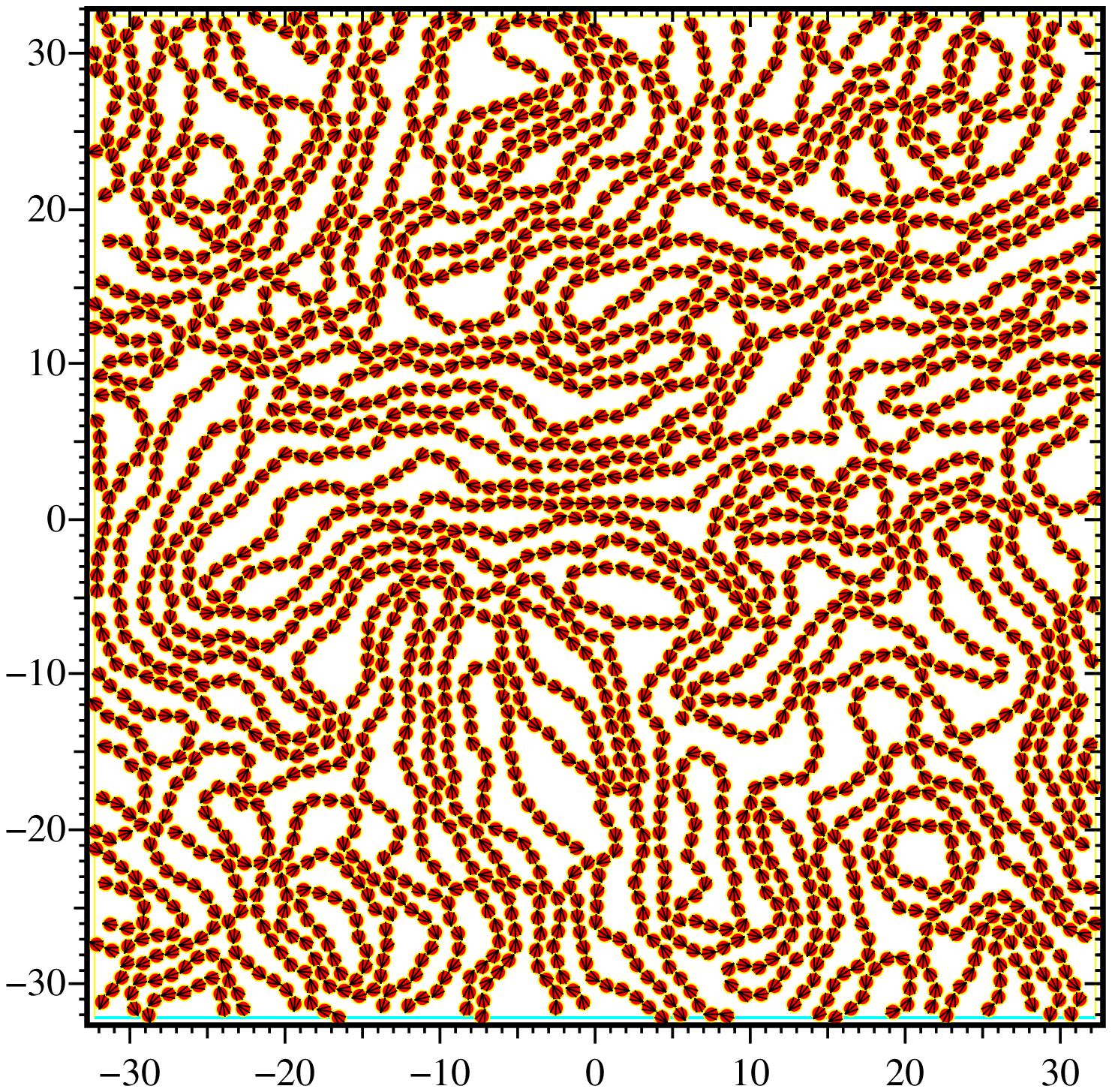} \\
\end{tabular}
\caption{Snapshots of configurations of the DHD system in space $\mathcal{C}_2$.  Top left : ($\rho^{*} =0.1$, $\mu^* = 2$, $N=1024$),
Top right : ($\rho^{*} =0.1$, $\mu^* = 3$, $N=1024$),
Bottom left : ($\rho^{*} =0.6$, $\mu^* = 2$, $N=1024$),
Bottom right : ($\rho^{*} =0.6$, $\mu^* = 3.5$, $N=2500$).
}
\label{Fig2}
\end{figure}

\begin{figure}[h!]
\centering
\begin{tabular}{ c  }
\includegraphics[scale=0.60]{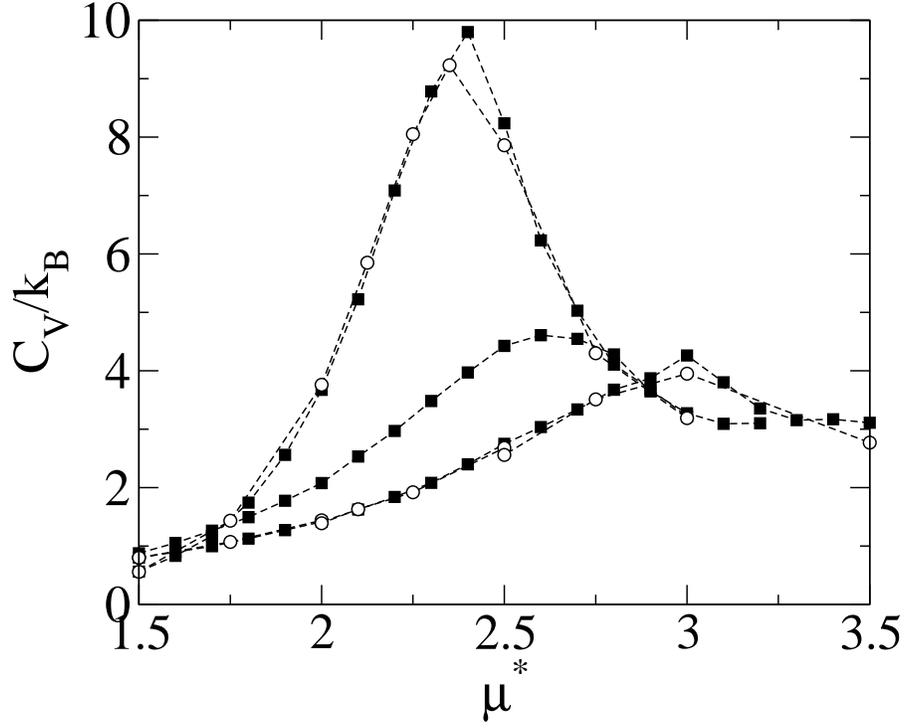} 
\end{tabular}
\caption{Specific heat $C_v/k_B$ versus reduced dipole moment $\mu^*$. Solid squares : $\mathcal{S}_2$, from
left to right $\rho^*=0.1, \, 0.4,\,  0.6$. Open circles :  $\mathcal{C}_2$, from
left to right $\rho^*=0.1, \,  0.6$. Dashed lines are guide lines for the eye.
}
\label{Fig3}
\end{figure}

%%%%%%%%%%%%%%%%%%%%%%%%%%%%%%%%%%%%%%%%%%%%%%%%%%%%%%%
 The homogeneous, isotropic fluid
phase is no more stable at low temperatures and complicated structures arise in this domain as indicated by some
snapshots displayed in Fig.~\ref{Fig2}. At low densities, clusters of aligned dipoles, mostly organized
into closed rings, 
appear at low temperatures and this topological
structure becomes even more complex at higher densities. In this low temperature regime the theory of the dielectric constant given
in Sec.~\eqref{dielectric} becomes incorrect and the predicted asymptotic behaviours of $h^{\Delta}(r)$, and $h^{D}(r)$
are no more observed. Most probably the dielectric tensor, even if  it exists, is no more isotropic and Fulton's theory
breaks down. In order to establish the thermodynamic stability of the high  temperature phase we have 
followed the authors of Ref.é\cite{workum:05,stamba:05} and   computed  the specific heat
$C_v$ as a function
of $\mu$ for some densities $\rho^*=0.05, 0.1, 0.2, 0.3, 0.4, 0.5, 0.6, \text{and } 0.7$.  A peak in $C_v(\mu)$
should be a signal of the ``transition'' or  the limit of stability of the fluid phase.
Some curves $C_v(\mu)$ are displayed in Fig.~\ref{Fig3}. They were obtained in the canonical ensemble
for systems involving $N\sim1000$ dipoles and runs of  $N_{\rm Conf.}=5-10 \times 10^6$ configurations
per particle.
Table~\ref{II} provides the transition dipole moments for the different densities considered. 
In Ref~\cite{workum:05} it has been pointed out that the
polymerization transition may also be defined from the inflection point of
$\Phi =N_p/N $ as a function of dipole moment (or temperature
$T^*= 1/\mu^{*2}$)  where $ N_p$ is the number of particles belonging
to a cluster.  At the density $\rho^*=0.05$ where clusters are well
defined we obtain a transition temperature in agreement with
 the value given in Table~\ref{II}.

%%%%%%%%%%%%%%%%%%%%%%%%%%%%%%%%%%%%%%
%%%%%%%%%%%%%%%%%%%%%%%%%%%%%%%%%%%%%%
\section{Conclusion}
\label{Conclu}
In this paper we have studied the  2D  DHD system by means of MC simulations
performed either in a square  with periodic boundary conditions or on the surface
of a sphere. The interactions between dipoles have been chosen so as to satisfy the laws
of electrostatics in the two geometries. With this precaution both methods lead to identical results for the thermodynamic, structural and
dielectric properties of the system, at least for sufficiently large systems. A subtlety  in the asymptotic
behaviours of the pair correlation function, strongly depending on the geometry, has been predicted and observed in the MC experiments
performed in the isotropic fluid phase.

In the low temperature, low density part of the phase diagram
a phase of living polymers of aligned dipoles organized into closed rings has been observed. At higher density the 
structure of this phase  looks like an entangled structure of chains and rings.

At these low temperatures the laws of macroscopic dielectrics
seem to be violated. A polymerization transition line based
on the maximum of the specific heat as a function of dipole
moment is provided. The critical dipole moment  $\mu_c^*$ at the
transition from fluid to polymeric phase increases slightly
with density.

We can contrast the present system with the one of  3D  dipolar particles with centers of mass 
constrained to a monolayer or thin layer, at least if the dipoles are in-plane as it is the case
at low temperatures. Such a quasi-two-dimensional (Q2D) system has been studied extensively in numerical
simulations~\cite{JJWeis_I,JJWeis_II,JJWeis_III,Kan_I,Kan_II} 
in view of its relevance to various experimental situations. References to experimental 
works can be found in~\cite{JJWeis_I,JJWeis_II,JJWeis_III,Kan_I,Kan_II}, see also
 Refs.~\cite{Butter_I,Butter_II,Klock,stamba:05,stamba:03}.

In Q2D systems the head to tail interaction of two particles at contact is $-2 \mu^2$ and
antiparallel side by side interaction is $-\mu^2$, while in  2D the interactions of both types of arrangement 
are of similar strength 
$-\mu^2$. One would therefore expect that chaining is much favoured in the Q2D case. 
This is easily  demonstrated by comparing structural properties obtained in
simulations of both systems. Notwithstanding, the overall qualitative structural
behaviour appears to be much the same at comparable densities (and short
range interaction), especially at low temperature, \textit{i.e.}, formation of
chains and rings.
A notable difference between the Q2D and 2D systems is however that in the
former system the spatial decay of the interaction is faster ($1/r^3$) than
the system dimension (D=2) \textit{i.e.}, of "short" range. Moreover,  the angular dependence
of the dipole-dipole interaction in  Q2D systems is a linear combination of 
the 2D  rotational invariants $D(1,2)$ and $\Delta(1,2)$. 

Although 2D dipolar
fluids do not exist \textit{per se} in nature, the model  could be used via various mappings for
applications as, recently, for the hydrodynamics of two-dimensional
microfluids of droplets. It is argued in Ref.~ \cite{Nature} that droplet velocities show
long-range orientational order decaying as $1/r^2$.

%%%%%%%%%%%%%%%%%%%%%%%%%%%%%%%%%%%%%%
%%%%%%%%%%%%%%%%%%%%%%%%%%%%%%%%%%%%%%

\begin{table}[h!]
\centering
\begin{tabular}{ c l l l l l l l l l }
                geometry & $\rho^*$        &       $\mu^* $   &     $\beta u$    &    $Z$         &     $\epsilon$   &  $g^{00}(\sigma)$  &    $h^{\Delta}(\sigma)$   & $h^{D }(\sigma)$ &    $C_V/k_B$      \\
\hline 
   $\mathcal{C}_2$&   $0.7 $      & $\sqrt{2}$ &     $ -1.789$   &   $3.990$   &    $17.45$        &   $ 4.346$             &        $2.826 $                 & $4.578 $       &   $.712  $  \\
\hline 
   $\mathcal{S}_2$&   $0.7 $      & $\sqrt{2}$ &     $ -1.790$   &   $3.974$   &    $17.83$        &   $ 4.339$             &        $2.814 $                 & $4.575$        &   $ .709 $ \\
\hline 
\hline 
  $\mathcal{C}_2$&   $0.6 $      & $2.0 $         &     $ -4.168$   &   $1.680$   &    $42.8$         &   $ 5.144$             &        $ 5.529$                 & $7.131 $       &   $ 1.44   $  \\
\hline 
  $\mathcal{S}_2$&   $0.6 $      & $2.0 $         &     $ -4.172$   &   $1.662$   &    $ 43.24$       &   $ 5.132$             &        $ 5.502$                 & $ 7.117$       &   $ 1.419   $  \\
\hline 
\hline 
   $\mathcal{C}_2$&   $0.1 $      & $2.0$ &     $ -1.957 $   &   $0.53$   &    $3.34$        &   $9.48  $             &        $8.03$                 & $ 15.91$       &   $ 3.76 $  \\
   $\mathcal{S}_2$&   $0.1 $      & $2.0$ &     $ -1.955 $   &   $0.53$   &    $3.32$        &   $9.45 $              &        $7.98$                 & $ 15.85$       &   $ 3.66 $  \\
\hline 
\hline 
\end{tabular}
\caption{\label{I}Geometry of simulation,  dimensionless numerical density $\rho^*$, reduced dipole moment $\mu^*$, reduced internal energy per particle  $\beta u$, 
dielectric constant $\epsilon$, compressibility factor $Z=\beta P / \rho$, contact values $g^{00}(\sigma)$,  $h^{\Delta}(\sigma)$, and $h^{D}(\sigma)$
of some  projections of the  pair correlation function,  and specific heat
of the DHD fluid. For each state systems of $N=1000$ particles were considered in  $\mathcal{S}_2$ and  $N=1024$ in  $\mathcal{C}_2$.
In both cases  $N_{\rm Conf.}= 5-10 \times 10^6$ configurations per particle were generated.}
\end{table}
%%%%%%%%%%%%%%%%%%%%%%%%%%%%%%%%%%%%%%
\newpage
\begin{table}[h!]
\centering
\begin{tabular}{ c c c c c c c c c  }
   $\rho^*$    &    $0.05$                  &     $0.1  $           &    $0.2  $            &     $0.3  $          &  $0.4 $          &   $0.5  $           &         $0.6$       \\
\hline 
   $\mu_c^* $&   $2. 35 \pm 0.15 $  & $2.4 \pm 0.1$ &  $ 2.5 \pm 0.1$  &  $2.5 \pm 0.25$ &  $2.6 \pm 0.1 $ &   $ 2.8 \pm 0.1$ &        $3.0 \pm 0.1$     \\
\hline 
\hline 
\end{tabular}
\caption{\label{II} $\mu^*_c(\rho^*)$ for a system of $N = 1000$ dipolar hard disks.}
\end{table}
%%%%%%%%%%%%%%%%%%%%%%%%%%%%%%%%%%%%%%
\newpage
%%%%%%%%%%%%%%%%%%%%%%%%%%%%%%%%%%%%%%%%%%%%%%%%%%%%%%%
\newpage
%%%%%%%%%%%%%%%%%%%%%%%%%%%%%%%%%%%%%%%%%%%%%%%%%%%%%%%
\newpage
%%%%%%%%%%%%%%%%%%%%%%%%%%%%%%%%%%%%%%%%%%%%%%%%%%%%%%%

%%%%%%%%%%%%%%%%%%%%%%%%%%%%%%%%%%%%%%
%%%%%%%%%%%%%%%%%%%%%%%%%%%%%%%%%%%%%%
%%%%%%%%%%%%%%%%%%%%%%%%%%%%%%%%%%%%%%

\end{document}